\documentstyle[12pt,a4]{article}
\begin{document}
\begin{flushright}
LMPM/95-15\par\noindent
hep-th/9510032
\end{flushright}

\title{Quantum Group Approach to a soluble vertex model 
with generalized ice-rule}

\author{ L. SOW CIRE\dag \hspace{0.15cm}\footnote{e-mail: sowcire@celfi.phys.univ-tours.fr} 
and T. T. TRUONG\ddag\footnote{e-mail: truong@u-cergy.fr}\\
\dag Laboratoire de Mod\`eles de Physique Math\'ematique,\\
D\'epartement de Physique, Universit\'e de Tours,\\
Parc de Grandmont, F-37200 Tours, France.\\
\ddag Groupe de Physique Statistique,\\
D\'epartement de Physique, Universit\'e de Cergy-Pontoise,\\
B.P 8428 F-95806 Cergy-Pontoise Cedex-France.}

\date{( to appear in Int. J. Mod. Phys. A (1995))}

\maketitle 
\begin{abstract}
\noindent
Using the representation of the quantum group $SL_q$(2) by 
the Weyl ope\-ra\-tors of the canonical commutation relations 
in quantum mechanics, we construct and solve a new vertex model 
on a square lattice. Random variables on horizontal bonds are 
Ising variables, and those on the vertical bonds take 
half positive integer values. The vertices is subjected to a 
genera\-li\-zed form of the so-called ``ice-rule'', its property 
are studied in details and its free energy calculated with the 
method of quantum inverse scattering. Remarkably in analogy with 
the usual six-vertex model, there exists a ``Free-Fermion'' limit with 
a novel rich operator structure. The existing algebraic structure 
suggests a possible connection with a lattice neutral plasma of charges, 
via the Fermion-Boson correspondence.\end{abstract}
\newpage
\section*{Introduction}

Vertex systems which are originally introduced to two dimensional 
ferroelectrics models with the so called ``ice-rule'', have rapidly 
evolved in the last quarter of century and have led to a tremendeous 
developement in the statistical physics of soluble models. 
New concepts and new mathematical structures have been discovered 
essentially due to the introduction of the method of quantum inverse 
scattering. Recently the solubility conditions known as the Yang-Baxter 
equations, which generalized the star-triangle relations for spin systems, 
for vertex systems are shown to have an expression in terms of quantum groups.
These new objects introduced by Drinfield \cite{drinf} and 
Jimbo \cite{jimbo} , are essentially matrix groups with non 
commutative elements. In this paper we shall be concerned with 
the $SL_q$(2) group which is a $q$-deformation of the standard $SL$(2) group \cite{bm}.\\

Interestingly, when one seeks a representation of the $SL_q$(2) 
group by the Weyl operators associated with a quantum degree 
of freedom $Q$ and its canonical conjugate momentum $P$, 
one obtains a new soluble vertex system having on ho\-ri\-zon\-tal 
bonds random variables taking only two values ($\pm$1/2), and 
on vertical bonds random variables taking an infinite number 
of values ( a kind of infinite spin ). Physically one may view 
the system as a two-dimensional array of vertical quantized spring 
coupled to binary horizontal devices. Alternatively the discrete
vertical variables are analogous to ``heights'' in face models
( or SOS models ) of statistical mechanics for which, there
exists a Yang-Baxter Algebra and Bethe-Ansatz solutions \cite{hv},
whereas the horizontal variables remain standard arrow variables
of Lieb's six-vertex model. In this respect, the new soluble model
may be viewed as having a mixed face-vertex nature.\\

Section 1 is devoted to the description of the system. We show how 
the construction of the vertex operator L is performed using 
$Q$ and $P$. We also indicated how a local vacuum may be chosen 
in order to be able to apply the method of quantum inverse scattering 
and to obtain the Bethe Ansatz equations.\\

In section 2, we diagonalized the transfer matrix of the model, 
using this conventional technique and obtain the free enegy per site. 
In analogy to the six-vertex model, we derive the tensor representation 
of the $SL_q$(2) group, the generators of which have a remarquable 
structure which is parallel to that of the six-vertex model. 
At the value $q$ = $i$, there is also a ``free fermion'' limit, 
and fermion-like operators may be expressed in terms of the Bose 
degrees of freedom $Q_{j}$ and $P_{j}$ for $j = 1, 2, \cdots, N.$ 
The generators of the tensor representation of $SL_q$(2) in this model, 
are closed to the ``screening'' operators of Dotsenko and Fateev \cite{dotko} 
in their treatment of Conformal Field theory using the Coulomb gas picture. 
If a Hamiltonian operator for a chain with degrees of freedom 
$Q_{j}$ and $P_{j}$ can be found so that it commutes with 
the transfer matrix of the model, one may be able to establish 
a connection with a lattice Coulomb gas. Since the critical behavior 
of these models are the same, there is a strong evidence that such 
a connection may exist. We conclude by comparing our model with 
those arising from the lattice version of the quantum Sine-Gordon  
or the quantum Non-linear Schr\"odinger equation 
( Faddeev , Korepin, Kulish, Sklyanin et al. \cite{fada}, \cite{kul}, \cite{kora}, 
and those arising from the bosonisation of the six-vertex 
model using the Holstein-Primakoff transformation 
(Y. K. Zhou \cite{zhou} ). Finally we give some 
future directions of investigation. 
\newpage
\section{The vertex system on a square lattice}
\subsection{Formulation}

The statistical system we consider is made up of elementary vertices 
consisting of a pair of Ising variables $ \sigma$ and $\sigma^{\prime}$ 
on horizontal bonds and other random variables on vertical bonds 
$\xi$ and $\xi^{\prime}$. For each set of values of the 4 random 
variables $\sigma$, $\sigma^{\prime}$ and $\xi$, $\xi^{'}$ 
a Boltzmann weight W$\xi,\xi^{'} ;\sigma ,\sigma^{'})$ is assigned (see fig.1).

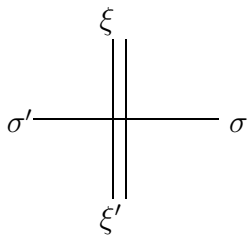
\begin{figure}[thp]
\begin{picture}(100,100)
\put(170,15){\line(0,1){60}}
\put(175,15){\line(0,1){60}}
\put(140,45){\line(1,0){70}}
\put(165,79){\shortstack{$\xi$}}
\put(165,4){\shortstack{$\xi'$}}
\put(130,40){\shortstack{$\sigma'$}}
\put(214,40){\shortstack{$\sigma$}}
\end{picture}
\caption[]{The elementary vertex with Boltzmann weight $W(\xi,\xi';\sigma,\sigma')$ .}
\label{Crossing}
\end{figure}

In the  1970's, R.J.Baxter \cite{baxa} showed that vertices 
that satify the triangle-relations (nowadays called the  
Yang-Baxter Equations) then their horizontal row transfer 
matrices form a commuting set of operators with respect 
to a ``spectral parameter''  introduced by the Russian school 
\cite{fada}. The triangle equations are the analog of the 
star-triangle relations for spin systems which yield then 
the same property of commuting horizontal row transfer matrices. 
The triangle-relations state that the partition function of the 
following two triangles are the same for every configuration of 
random variables on open external bonds (i.e. 
$\sigma_1$,$\sigma_2$,$\sigma_3$,$\sigma_4$,$\xi_1$,$\xi_2$) (see fig.2).\\
\begin{figure}[thp]
\unitlength=1cm
\begin{picture}(10,4.5)
\put(0.8,3){\line(5 $\sigma_4$ , -1){4.5}}
\put(0.75,2.5){\line(5 $\sigma_3$ , 1){4.5}}
\put(4.7,4){\line(0,-1){2.5}}
\put(4.5,4){\line(0 ,-1){2.5}}
\put(7.5,3.3){\line(5 $\sigma_4$ , -1){4.5}}
\put(8,2){\line(5,1){4.5}}
\put(8.9,4){\line(0,-1){2.5}}
\put(8.7,4){\line(0,-1){2.5}}
\put(5.9,3.5){\shortstack{$\sigma_1$}}
\put(7.5,2){\shortstack{$\sigma_3$}}
\put(8.5,1.2){\shortstack{$\xi_2$}}
\put(4.55,1.2){\shortstack{$\xi_2$}}
\put(4.9,2.5){\shortstack{$\xi$}}
\put(8.2,2.5){\shortstack{$\xi$}}
\put(4.55,4.3){\shortstack{$\xi_1$}}
\put(8.5,4.3){\shortstack{$\xi_1$}}
\put(12.8,3){\shortstack{$\sigma_1$}}
\put(6,2){\shortstack{$\sigma_2$}}
\put(12.8,2.3){\shortstack{$\sigma_2$}}
\put(6.6,2.6){\shortstack{=}}
\put(10,2){\shortstack{$\sigma$}}
\put(10,3.1){\shortstack{$\sigma'$}}
\put(3.5,2){\shortstack{$\sigma'$}}
\put(3.5,3.1){\shortstack{$\sigma$}}
\end{picture}
\label{crossing}
\caption[]{Summation is performed on $\sigma,\; \sigma'$  and, $\xi$}
\end{figure}
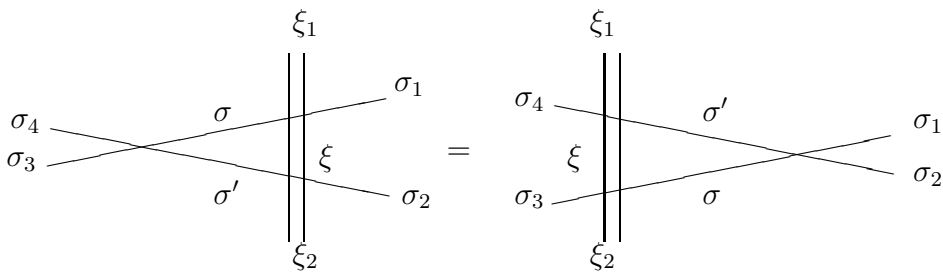
\newpage
Here as it stands, we note the necessity of having a third vertex 
having only Ising variables on the left and right sides.
The triangle-relations are the necessary conditions for 
the calculation of the partition function of the model 
by Bethe Ansatz techniques and consequently the thermodynamics 
of the system. Note that when $\xi=\sigma=\pm1$, we recover the 
standard triangle relations of the six or eight vertex solved by 
E. H. Lieb \cite{lieb} and R. J. Baxter \cite{baxa}. A particular 
system of vertex having unequal number of random variables on 
horizontal and vertical bonds but verifying the triangle relations 
was solved by R. Z. Bariev and Yu. V. Kozhinov \cite{bara}, and a 
general discussion on such type of vertex is presented by H. J. de Vega \cite{vega} \\
\indent
An appropriate way of handling the star triangle relations 
(or Yang-Baxter equations) consists of using an operator formulation.
We associate to a vertex with Ising variables on horizontal bonds 
a 2$\times$2 matrix whose matrix elements $L_{\sigma\sigma^{'}}$ 
are labelled by $\sigma$ and $\sigma^{\prime}$ (see fig.1):
\begin{displaymath}
L=\left(\begin{array}{ccccccc}L_{11} & = & \alpha & \hspace{1cm} & L_{1-1} & = &\beta^-\\
     L_{-11} & = & \beta^{+} & \hspace{1cm} & L_{-1-1} & = & \delta
\end{array} \right)
\end{displaymath}\\
$\alpha$, $\beta^\pm$, $\delta$ are themselves operators in a ``vertical'' 
Hilbert space with matrix elements labelled by $\xi$ and $\xi^{'}$. 
Anticipating on the existence of a ``spectral parameter'' \cite{fadb}, 
the triangle-relations take up the following compact form:\\
\begin{equation}
R(\frac{u}{v})L(u)\otimes L(v)=L(v)\otimes L(u)R(\frac{u}{v}),
\end{equation}\\
where $R$ is associated with the standard vertex with only Ising variables on the bonds.\\

Eq(1) is rather general, one may consider the set of all $L$ 
satisfying (1) with the same $R$; in particular one may choose 
$R$ to be that of a standard symmetric six-vertex model. 
In this case the matrix elements $\alpha$(u), $\delta$(u), $\beta^\pm$ 
of $L$ are shown to be expressible in terms of the generators 
of the quantum group $SL_q$(2): $L_z$, $L^{+}$, $L^{-}$ which 
obey the defining relations (see \cite{wieg}):\\
$$[L_z,L^\pm ] = \pm L^\pm \hspace{0.5cm}{\mbox or}\hspace{0.5cm}
q^{L_{z}}L^{\pm} = q^{\pm1}L^{\pm}q^{L_{z}},$$
\begin{equation}
[ L^+ ,L^- ] = \frac{q^{2L_z} - q^{-2L_z}}{q - q^{-1}}.
\end{equation}

Then one has the expressions ( see Wiegmann and Zabrodin \cite{wieg} )                    
\begin{equation}                                                             
\beta^{\pm} = L^{\pm} ,
\end{equation}                                                                 
$$\alpha(u) = \frac{uq^{L_z} - u^{-1}q^{-L_z}}{q - q^{-1}} , $$ 
\begin{equation}
\delta(u) = \frac{uq^{-L_z} - u^{-1}q^{L_z}}{q - q^{-1}} .
\end{equation}
Thus the problem is reduced to finding the appropriate  representations 
of $SL_q(2)$ which corresponds to the definition of the vertex. 
We observe that the standard six vertex model is recovered if one 
considers the spin 1/2 representation of $SL_q(2)$ which is 
generated by the Pauli matrices, higher spin representations 
are possible (see Saleur and Pasquier \cite{sala}, \cite{pas} ).\\

But elaborating on Wiegmann and Zabrodin \cite{wieg}  representation 
of $SL_q(2)$ by 2-dimensional magnetic translation operators, we shall 
consider next, the representation of $SL_q(2)$ by  the Weyl operators 
of the canonical commutation relation of one degree of freedom in quantum mechanics.
\subsection{Algebraic tools}

In his treatement of quantum mechanics H. Weyl had proposed to replace 
the canonical commutator between a dynamical variable $Q$ and its conjugate $P$\\
\begin{equation}
[Q,P]=iI ,
\end{equation}\\
by the commutation relation between $e^{ipQ}$ and $e^{ixP}$:\\
\begin{equation}                                                                
e^{ixP}e^{ipQ} = e^{ixp}e^{ipQ}e^{ixP} .
\end{equation} \\

 In this paper we shall concentrate on the case $q=e^{i\eta}$, 
which corresponds to a physical phase of the system. 
Using appropriate scaling, (6) may be rewritten under the form:\\
\begin{equation}\\
e^{iP}q^Q = qq^Qe^{iP} .
\end{equation} \\
This relation allows us to represent the generators of $SL_q(2)$ as:\\
$$L_z = Q , $$
$$L^+ = \frac{q^{Q-1/2}-q^{-Q+1/2}}{q-q^{-1}}e^{-iP} ,$$
\begin{equation}
L^- = -e^{iP}\frac{q^{Q-1/2}-q^{-Q+1/2}}{q-q^{-1}} .
\end{equation} \\
From (8), we recover in the limit q$\longrightarrow$1, the 
following representation of $SL(2)$ generators:
$$S^+ = (Q-1/2)\exp(-iP),$$
$$S^- = -\exp(iP)(Q-1/2),$$
$$S^z = Q.$$
We observe that $L^{+}$ and $L^{-}$ are each other antihermitian 
when we require that the limit $q\longrightarrow 1$ of (8)
obeys the commutators of $SL$(2).\\

One may check that (8) fulfills automatically (2), using the shift property of $e^{ixP}$,\\
\begin{equation}
e^{ixP}Qe^{-ixP} = Q+x .
\end{equation}\\
The new representation of the vertex operator $L$ is now:
$$\beta^{+} = \frac{q^{Q-1/2}-q^{-Q+1/2}}{q-q^{-1}}e^{-iP} ,\hspace{1cm}
\beta^{-} = -e^{iP}\frac{q^{Q-1/2}-q^{-Q+1/2}}{q-q^{-1}} ,$$

\begin{equation} 
\alpha = \frac{uq^{Q}-u^{-1}q^{-Q}}{q-q^{-1}},\hspace{1cm}                                                             
 \delta = \frac{uq^{-Q}-u^{-1}q^{Q}}{q-q^{-1}} .
\end{equation}\\                                                                
Physically we have a quantum mechanical degree of freedom on ``vertical'' 
space coupled to Ising spins on horizontal bonds. Such a system obeys 
the triangle-relations(Yang-Baxter equations) and may under specified 
conditions be solved by Bethe Ansatz techniques. Since it has the $R$-matrix 
of a six-vertex model, one expects its critical behavior to be the same 
as in the six-vertex case. The attractive point is that the critical 
universality class may be in fact defined by the choice of the $R$-matrix.\\

For comparison let us recall the standard spin 1/2 representation 
of $SL_q(2)$, for which we have: 
$$\frac{q^{\sigma^z}-q^{-\sigma^z}}{q-q^{-1}} =\sigma^z .$$\\
In this case with the standard Pauli matrices $\sigma^x$, $\sigma^y$, $\sigma^z$, one has:\\
$$\beta^\pm = \sigma^\pm =1/2(\sigma^x \pm i\sigma^y) ,$$
$$\alpha = \frac{uq^{\sigma^{z}/2} - u^{-1}q^{-\sigma^{z}/2}}{q - q^{-1}} ,$$
\begin{equation}                                                               
\delta = \frac{uq^{-\sigma^{z}/2} - u^{-1}q^{\sigma^{z}/2}}{q - q^{-1}} .
\end{equation} \\
Here the vertical space is two-dimensional, whereas it is infinite dimensional 
when one uses $P$ and $Q$.\\
\subsection{Schr\"odinger representation and the local vacuum.}

In order to apply the method of quantum inverse scattering 
(Faddeev \cite{fadb}) to construct the explicit solution of the problem,
one needs to construct a local vacuum. This is fairly evident 
in the case of the standard six-vertex model where one may choose 
for example, the state 
$\left(\begin{array}{c}0\\1\end{array}\right)$ which is
annihilated by the $\beta^-$ = $\sigma^-$ operator:
\begin{equation}
\sigma^{-}\left(\begin{array}{c}0\\1\end{array}\right) = 0 .
\end{equation}
Thus the local vacuum is nothing else as the ``spin down'' state on 
the vertical direction. Moreover\\
$$\alpha\left(\begin{array}{c}0\\1\end{array}\right) = 
\frac{uq^{-1/2} - u^{-1}q^{1/2}}{q - q^{-1}} 
\left(\begin{array}{c}0\\1\end{array}\right) = 
a \left(\begin{array}{c}0\\1\end{array}\right) ,$$
\begin{equation}
\delta \left(\begin{array}{c}0\\1\end{array}\right) =
\frac{uq^{1/2} - u^{-1}q^{-1/2}}{q - q^{-1}}\left(
\begin{array}{c}0\\1\end{array}\right) = b \left(
\begin{array}{c}0\\1\end{array}\right) .
\end{equation}\\ Or in a more usual parametrization : \,
$q = e^{i\eta } ,\; \; u = e^{\theta} ;$
we recognize:
$$a = \frac{\sinh{(\theta - i\eta/2)}}{\sinh{i\eta}} ,
\hspace{2cm} b = \frac{\sinh{(\theta + i\eta/2)}}{\sinh{i\eta}} .$$
\indent
With the use of $Q$ and $P$ it is necessary to find a 
local vacuum. We shall do so in using the position 
Schr\"odinger representation, in which the $Q$ operator 
is diagonal and has continuous spectrum:\\
$$Q \vert\xi\rangle = \xi\vert\xi\rangle ,$$
\begin{equation}
e^{ixP} \vert\xi\rangle = \vert\xi - x\rangle .
\end{equation}
The local vacuum $|\omega\rangle$, in analogy with eq( 12 ), 
is defined by the annihilation property \\
$$\beta^{-} \vert\omega\rangle = L^{-} \vert\omega\rangle = 0 $$
 \begin{equation}  
e^{iP}\frac{q^{{\omega}-{1/2}} - q^{-{\omega}+1/2}}{q - q^{-1}} \vert\omega\rangle = 0 .
\end{equation}
Hence one must choose 
\begin{equation}
\omega = 1/2 .
\end{equation}
\indent
The existence of this local vacuum is directly related to 
the construction of the representation
(see eq.(8)) with a proper limit $q\longrightarrow1$. 
Demanding from the start that $L^{+}$ and
$L^{-}$ be each other hermitian, will not lead to the 
correct $q\longrightarrow1$limit, nor yield
a local vacuum for one vertex operator $L$, as in the 
Sine-Gordon theory \cite{kora}.    
Application of $\beta^+ = L^{+}$ on the local vacuum $\vert \omega\rangle = 
\vert1/2\rangle$ yields\\
$$L^{+}\vert 1/2\rangle = \vert 3/2\rangle .$$\\
More generally we have:\\
\begin{equation}
(L^{+})^n \vert 1/2\rangle = \frac{q^n -q^{-n}}{q - q^{-1}} \times 
\frac{q^{n-1} - q^{-n+1}}{q - q^{-1}} \times \ldots \times 
\frac{q - q^{-1}}{q - q^{-1}} \vert n+1/2\rangle .
\end{equation}\\
The sequence of states $\vert n+1/2\rangle$ reminds us of 
the sequence of the harmonic oscillator with unit frequency. 
In this sense it is reasonnable to think of the vertices as vertical 
springs coupled to horizontal Ising variables.\\

We may evaluate also, since $Q$ is diagonal, the action of 
$\alpha(u)$ and $\delta(u)$ on $\vert 1/2\rangle$ \\
$$\alpha ( u ) \vert 1/2\rangle =\frac{uq^{1/2} - 
u^{-1}q^{-1/2}}{q - q^{-1}} \vert 1/2\rangle ,$$
\begin{equation}
\delta ( u ) \vert 1/2\rangle = \frac{uq^{-1/2} - 
u^{-1}q^{1/2}}{q - q^{-1}} \vert 1/2\rangle .
\end{equation}
This will be used later in constructing the solution.

We can check that the set of states $\vert n+1/2\rangle$ 
generated by application of $(L^{+})^n$ on $\vert 1/2\rangle$, 
the local vacuum, forms an orthonormal set of states for the ``vertical'' Hilbert space.
The proof is as followed:
from (2) we have, \\
\begin{equation}
[L^{-} ,(L^{+})^{m}] = -m(L^{+})^{m-1}\frac{q^{2(Q+m-1)} - q^{-2(Q+m-1)}}{q - q^{-1}}.
\end{equation}
Thus$$\langle n + 1/2\vert m + 1/2\rangle = \langle 1/2\vert 
(L^{-})^{n}(L^{+})^{m}\vert 1/2\rangle.$$
Using the former equation, we arrive at:
$$\langle n + 1/2\vert m + 1/2\rangle = \prod_{j=m-n+1}^{m}s_{j} 
\langle 1/2\vert (L^{-})^{m-n}\vert 1/2\rangle,$$
where$$ s_{j} = -j\frac{q^{(1/2 + j - 1)} - q^{-(1/2 + j - 1)}}{q - q^{-1}}.$$
With an adequate normalization, we have:
\begin{equation}   
\langle n + 1/2\vert m + 1/2\rangle = \delta_{mn}.
\end{equation}

Hence we can label the states on the vertical bonds of the vertices 
by half integer $(n+1/2)$, wich are similar to ``height variables''
in SOS models, or face variables \cite{hv}. 
We get four families of vertices with corresponding 
Boltzmann weights with $\sigma$, $\sigma'$ = $\pm$1/2 on horizontal bonds:
$\vspace{0.5cm}$
\begin{figure}[thp]
\begin{picture}(100,120)    
\put(90,45){\line(0,1){60}}
\put(95,45){\line(0,1){60}}
\put(60,75){\line(1,0){70}}
\put(60,75){\vector(1,0){15}}
\put(60,75){\vector(1,0){60}}

\put(320,75){\vector(-1,0){15}}
\put(250,75){\vector(1,0){15}}
\multiput(280,45)(0,0){1}{\line(0,1){60}}
\multiput(285,45)(0,0){1}{\line(0,1){60}}
\multiput(250,75)(0,1){1}{\line(1,0){70}}

\put(98,105){\shortstack{$n$+1/2}}
\put(98,45){\shortstack{$n$+1/2}}
\put(288,105){\shortstack{$n$+3/2}}
\put(288,45){\shortstack{$n$+1/2}}

\put(11,75){\shortstack{$\sigma'$=+1/2}}
\put(137,75){\shortstack{$\sigma$=+1/2}}
\put(208,75){\shortstack{$\sigma'$=1/2}}
\put(327,75){\shortstack{$\sigma$=-1/2}}

\put(252,20){\shortstack{$\displaystyle\omega_{{1\over2}{-{1\over2}}} = -\frac{q^{n+1} - q^{-(n+1)}}{q - q^{-1}}$}}
\put(68,20){\shortstack{$\displaystyle\omega_{{1\over2}{1\over2}} = \frac{uq^{n+1/2} -u^{-1}q^{-(n+1/2)}}{q - q^{-1}}$}}

\end{picture}
\begin{picture}(100,120)    

\put(90,45){\line(0,1){60}}
\put(95,45){\line(0,1){60}}
\put(130,75){\vector(-1,0){60}}
\put(60,75){\vector(1,0){60}}

\put(320,75){\vector(-1,0){60}}
\put(320,75){\vector(-1,0){15}}

\multiput(280,45)(0,0){1}{\line(0,1){60}}
\multiput(285,45)(0,0){1}{\line(0,1){60}}
\multiput(250,75)(0,1){1}{\line(1,0){70}}

\put(98,45){\shortstack{$n$+1/2}}
\put(98,105){\shortstack{$n$-1/2}}
\put(11,75){\shortstack{$\sigma'$=-1/2}}
\put(137,75){\shortstack{$\sigma$=+1/2}}

\put(62,20){\shortstack{$\displaystyle\omega_{-{1\over2}{1\over2}} = \frac{q^{n} - q^{-n}}{q - q^{-1}}$}} 
\put(246,20){\shortstack{$\displaystyle\omega_{-{1\over2}{-{1\over2}}} = \frac{uq^{-(n+1/2)} - u^{-1} q^{n+1/2}}{q - q^{-1}}$}} 

\put(288,45){\shortstack{$n$+1/2}}
\put(288,105){\shortstack{$n$+1/2}}
\put(206,75){\shortstack{$\sigma'$=-1/2}}
\put(327,75){\shortstack{$\sigma$=-1/2}}

\end{picture}
\caption[]{The four families of vertices with Boltzmann weights $\omega_{\sigma'\sigma}$.}
\label{Crossing}
\end{figure}

 We observe that in this case we still have an ``ice rule'' for the vertices (see fig.4)\\
\begin{figure}[thp]
\begin{picture}(100,85)
\put(170,15){\line(0,1){60}}
\put(175,15){\line(0,1){60}}
\put(140,45){\line(1,0){70}}
\put(165,79){\shortstack{$n$+1/2}}
\put(165,4){\shortstack{$n'$+1/2}}
\put(130,40){\shortstack{$\sigma'$}} 
\put(214,40){\shortstack{$\sigma$}}
\end{picture}
\caption[]{The ``ice-rule'' is of the form: $n+\sigma = n^{'} + \sigma^{'}$.}
\label{crossing}
\end{figure}
\paragraph*{}
This is the reason why the model is closed to the six-vertex and 
may be considered as an extension of the ``anisotropic'' vertices 
considered by Bariev \cite{bara}, where one of the ``vertical'' random 
variable may take infinite number of~ va\-lues~.~ The~vertical~operators 
$\alpha, \delta, \beta^{\pm}$ may be then represented in this base by 
infinite dimensional matrices and one may consider discussing the rotated 
representation (see Vega) \cite{vega}.

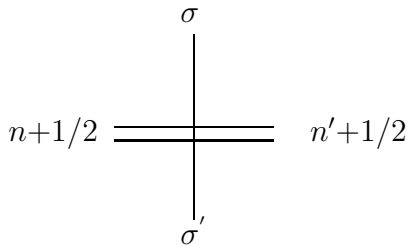
\begin{figure}[thp]
\begin{picture}(100,100)
\put(170,15){\line(0,1){70}}
\put(140,50){\line(1,0){60}}
\put(140,45){\line(1,0){60}}
\put(165,90){\shortstack{$\sigma$}}
\put(165,6){\shortstack{$\sigma^{'}$}}
\put(100,45){\shortstack{$n$+1/2}} 
\put(214,45){\shortstack{$n'$+1/2}}
\end{picture}
\caption{The rotated representation of the vertices.}
\label{crossing}
\end{figure} 
Because of the ``ice-rule'', the weights $\omega_{{1\over 2}{-{1\over 2}}}$
and $\omega_{-{1\over 2}{1\over 2}}$ will appear always in pairs.
We may then take out a factor $i$ and redefine the ``spectral parameter''
as $u' = -iu$, with $u'$ real, to obtain:
$$\omega_{{1\over 2}{1\over 2}} = \frac{u'q^{n+1/2} + (u')^{-1}q^{-(n+1/2)}}{q - q^{-1}}
\hspace{1.2cm}\omega_{{1\over 2}{-{1\over 2}}} = i\frac{q^{n+1} - q^{-(n+1)}}{q - q^{-1}},$$

\begin{equation}
\omega_{{-{1\over 2}}{1\over 2}} = -i\frac{q^{n} - q^{-n}}{q - q^{-1}}
\hspace{1.2cm}\omega_{-{1\over 2}{-{1\over 2}}} = \frac{u'q^{-(n+1/2)} + 
(u')^{-1}q^{n+1/2}}{q - q^{-1}}.
\end{equation}

Then the partition function will appear as positive definite. 
This unphysical form of weights arises often in the derivation 
of the parametrisation of the weights using the Yang-Baxter 
equations (see for example the case of the eight-vertex model \cite{bax71}).

Finally we observe that the Casimir operator of the $SL_{q}$(2) 
representation is independent of $q$, and has the value 0.
This may be checked by direct calculation (see \cite{wieg}):
$${\it C} = (\frac{q^{-1/2}q^{Q} - q^{1/2}q^{-Q}}{q - q^{-1}})^{2} + \beta^{+}\beta^{-} = 0.$$

\section{Properties of the vertex system}
\subsection{Quantum inverse scattering and diagonalisation of the transfer matrix}

Following standard procedure we construct the monodromy operator $T(u)$, 
for one row of $N$ vertices (see fig.5).\\
\begin{figure}[thp]
\begin{picture}(100,75)
\put(90,15){\line(0,1){60}}
\put(95,15){\line(0,1){60}}
\put(60,45){\line(1,0){150}$\hrulefill\ldots\hrulefill$}
\put(225,45){\line(1,0){145}}

\multiput(120,15)(210,0){2}{\line(0,1){60}}
\multiput(125,15)(210,0){2}{\line(0,1){60}}

\multiput(150,15)(150,0){2}{\line(0,1){60}}
\multiput(155,15)(150,0){2}{\line(0,1){60}}

\put(305,4){\shortstack{$L_{N-1}$}}
\put(145,4){\shortstack{$L_{3}$}}
\put(335,4){\shortstack{$L_{N}$}}
\put(85,4){\shortstack{$L_{1}$}}
\put(115,4){\shortstack{$L_{2}$}}
\end{picture}
\caption[]{A row of vertices.} 
\label{crossing}
\end{figure}
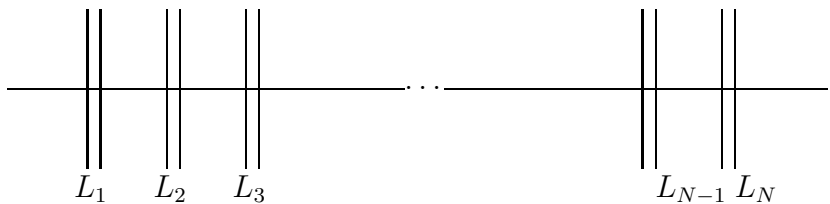

\paragraph*{}
As it is known, $T(u)$ is a $2\times 2$ matrix of the type,
\begin{equation}
T(u) = \left(
\begin{array}{ccc}
A(u)& &C(u)\\B(u)& &D(u)
\end{array}
\right)
\end{equation}
and obeys the triangle-equations (Yang-Baxter equations) 
which yield commutation relations between $A(u), B(u), C(u)$ and $D(u)$, 
necessary to construct the eigenstates of the transfer matrix
\begin{equation}
t(u) = \mbox{Tr}\;T(u) = A(u) + D(u).
\end{equation}

But before doing so, let us recall that due to our choice of local vacuum 
$\vert 1/2\rangle$ at each site, the local vertex system fulfills an extended 
``ice-rule'' on which, conservation laws relie on. The bare vacuum of the row is thus:
\begin{equation}
\vert \Omega \rangle = \vert 1/2 \rangle_{1} \otimes \vert 1/2 \rangle_{2} 
\otimes \ldots \otimes \vert 1/2 \rangle_{N}.
\end{equation}
Excitations are created in a standard way by repeated applications of 
the $B(u)$ operator on $\vert \Omega \rangle$, namely:
\begin{equation}
B(u_{1})B(u_{2})\ldots B(u_{m}) \vert \Omega \rangle = \vert u_{1},u_{2},
\ldots ,u_{m} \rangle.
\end{equation} 

The requirement that $\vert u_{1},u_{2},\ldots ,u_{m}\rangle$ should be 
eigenstate of $t(u)$ forces the $u_{j}, j = 1,2,\ldots, m$, to obey a system 
of nonlinear equations, the so-called Bethe Ansatz equations:\\
\begin{equation}
\frac{\sinh({\theta_{j} + i{\eta\over2}})}{\sinh({\theta_{j} - i{\eta\over2}})}
= -\prod_{k=1}^{m}\frac{\sinh({\theta_{j} - \theta_{k} + i\eta})}{\sinh({\theta_{j} -\theta_{k} - i\eta})},
\end{equation}
where $u_{j} = \exp{\theta_{j}}$.\\

These equations may be solved exactly in the limit of $N\longrightarrow \infty$ 
by standard Fourier techniques since they are formaly the same equations 
as in the six-vertex model \cite{vega}. The free energy per site is then 
in the thermodynamics limit in the antiferroelectric regime for example,\cite{vega}
\begin{equation}
f(\theta,\eta) \approx \int\limits_{0}^{+\infty}\frac{\sinh{2x\theta}
\sinh{[x(\pi-\eta)]}}{\cosh(x\eta)\sinh(x\pi)}\;dx + const,
\end{equation} 
and has the same critical behavior as the six-vertex or model I of 
\cite{hv}. This is not new since 
the lattice version of the Sine-Gordon model (Faddeev et al. \cite{fadb}, 
Korepin et al. \cite{kora}) shares also the same behavior. One may say that 
so long one has the same $R$-matrix in the Yang-Baxter equations, one necessarily 
obtain the same critical behavior and conjecture that, a possible classification 
of universality classes of critical behavior, may be made according to a classification of $R$-matrices.  
\subsection{Alternative Quantum Group structure and representation}
\indent
In this section we discuss some aspects of $SL_q$(2) which are relevant 
for the so called ``Free Fermion'' limit: q$\longrightarrow i$ 
(or $\eta \longrightarrow {\pi\over 2}$), which is well known in 
the Bethe Ansatz equations (26) (see for example \cite{baxlivre}).\\

Consider the representation of one vertex operator $L$ in term of 
the generators $L_{z}$, $L^{\pm}$ of $SL_q$(2). We may form two 
new operators $a^{+}$ and $a^{-}$ by defining:
\begin{equation}
a^{\pm} = q^{L_{z}}L^{\pm}.
\end{equation}
Then $SL_{q}$(2) may be characterized by $L_{z}$, $a^{\pm}$ 
with the following relations:
$$q^{L_{z}}a^{\pm} = q^{\pm1}a^{\pm}q^{L_{z}},$$
\begin{equation}
\displaystyle
a^{+}a^{-} - q^{-2}a^{-}a^{+} = \frac{q^{4L_{z}} - 1}{q^{2} - 1}.
\end{equation}

Curiously this last equation appears as a $q$-commutator for a 
$q$-deformed oscillator, although not identical to the relation 
proposed by Biedhenharn and Macfarlane \cite{bm}. The limit $q\longrightarrow 1$ 
of (29) is again $SL$(2) and the oscillator algebra can be recovered after a 
Wigner contraction is performed \cite{gil}.\\ 
 Moreover we see that for any integer $m$,
\begin{equation}
\displaystyle
(a^{+})^{m} = q^{-m(m-1)\over 2}q^{mL_{z}}(L^{+})^{m}.
\end{equation}
There exists a similar equation for $(a^{-})^{m}$. Note that $a^{+}$ and 
$a^{-}$ are not each other hermitian.\\

As it is known \cite{vega}, one may obtain tensor representation of 
$SL_q$(2) by repeated application of the coproduct (\cite{jimbo}) 
on the generators for one vertex. This is what one obtains physically 
when the whole row of N vertices is considered. The N-fold tensor representation 
generators are:
$$J_{\pm} = \sum_{j=1}^{N}q^{(L_{z})_{1}}\otimes \cdots \otimes q^{(L_{z})_{j-1}}
\otimes L_{j}^{\pm}\otimes q^{-(L_{z})_{j+1}}\otimes \cdots q^{-(L_{z})_{N}},$$
\begin{equation}
J_{z} = \sum_{j=1}^{N} (L_{z})_{j},
\end{equation}
they fulfill the defining relations (2):
$$[J_{+}, J_{-}] = \frac{q^{2J_{z}} - q^{-2J_{z}}}{q - q^{-1}},$$
\begin{equation}
q^{J_{z}}J_{\pm} = q^{\pm1}J_{\pm}q^{J_{z}}.
\end{equation}

These relations may be obtained by hand starting from the standard commutation 
relations of $A(u)$, $B(u)$, $C(u)$ and $D(u)$, obtained in the quantum 
inverse scattering method, in the ferroelectric regime ($q$ real and positive), 
and taking the appropriate limits \cite{vega}.\\

Similarly the tensor representation may be characterized
by the set of operators $J_{z}$, $\eta^{\pm}$
where,
\begin{equation}
\eta^{\pm} = q^{J_{z}}J_{\pm} = \sum_{j=1}^{N}q^{2(L_{z})_{1} + \cdots + 
2(L_{z})_{j-1}}a_{j}^{\pm}.
\end{equation}
Of course $\eta^{\pm}$, $J_{z}$ obey the relation (29) satisfied by $a^{\pm}$, 
$L_{z}$. But here, we have the freedom
of introducing new local operators $\eta_{j}^{\pm}$ 
defined by: 
\begin{equation}
\displaystyle
\eta_{j}^{\pm} = q^{2(L_{z})_{1} + \cdots + 2(L_{z})_{j-1}}a_{j}^{\pm},
\end{equation}
which in turns obey the following commutation relations:

$$\eta_{j}^{+}\eta_{i}^{+}  =  q^{2sgn(j-i)}\eta_{i}^{+}\eta_{j}^{+},$$
$$\eta_{j}^{-}\eta_{i}^{-}  =  q^{-2sgn(j-i)}\eta_{i}^{-}\eta_{j}^{-},$$
$$\eta_{j}^{+}\eta_{i}^{-}  =  q^{-2}\eta_{i}^{-}\eta_{j}^{+}\;\;i\not=j,$$
and
\begin{equation}
\eta_{j}^{+}\eta_{j}^{-} - q^{-2}\eta_{j}^{-}\eta_{j}^{+} = q^{4(L_{z})_{1} + 
\cdots + 4(L_{z})_{j-1}}\frac{q^{4(L_{z})_{j}} - 1}{q^{2} - 1}.
\end{equation}

Note that $(\eta_{j}^{+})^{m}$ or  $(\eta_{j}^{-})^{m}$ for $m$ integer, is not necessarily zero, 
because of (30).\\

Consequently in this section, we have shown that the structure of $SL_{q}$(2), 
presents aspects of the so called
$q$-deformed oscillator of a special type and that, its tensor representation 
leads naturally to local operators
which exhibit ``anyonic'' commutation relations as well as behaves the $q$-deformed 
oscillator \cite{bm}, \cite{wz}.  
\subsection{Free Fermion limit}

As can be seen from eq (26), the Bethe Ansatz equations decouple at
$\eta = {\pi\over 2}$ (or $q = i$): the scattering of pseudo-particles created by 
the $B(u)$ operators becomes trivial, the corresponding phase shifts
 equal -1, and the multiparticle wave function is simply represented by a Slater determinant.\\

We can now make some interesting observations in two different situations:
the six-vertex case and the vertex studied in this paper.
\subsubsection{The six-vertex case}

The $\eta^{\pm}$ operators are closely related to usual lattice Fermion operators. 
Since $L_{z} = {1\over 2}\sigma^{z},$
\begin{equation}
\eta_{j}^{\pm} = e^{i{\pi\over 2}(\sigma_{1}^{z} + \cdots + \sigma_{j-1}^{z})}e^{i{\pi\over 4}
\sigma_{j}^{z}}\sigma_{j}^{\pm},
\end{equation}
and they anticommute:

$$\eta_{j}^{\pm}\eta_{i}^{\pm} = -\eta_{i}^{\pm}\eta_{j}^{\pm},$$
\begin{equation}
\{\eta_{j}^{+}, \eta_{i}^{-}\} = \delta_{ij}e^{i\pi(\sigma_{1}^{z} + \cdots + 
\sigma_{j-1}^{z})}\frac{e^{i\pi\sigma_{j}^{z}} - 1}{-2}
= \delta_{ij} (-1)^{j-1}.
\end{equation}

Moreover $(\eta_{j}^{\pm})^{2} = 0.$ Thus we may define Fermion operators by:
\begin{equation}
\displaystyle
c_{j}^{\pm} = (i)^{j-1}\eta_{j}^{\pm} = \displaystyle \exp\{{i{\pi\over 2}(\displaystyle
\sum_{l=1}^{j-1}(\sigma^{z} +1)_{l})}\}\displaystyle 
\exp\{{i{\pi\over 4}\sigma_{j}^{z}}\}\sigma_{j}^{\pm},
\end{equation}
which are not exactly identical to the ones obtained by the Jordan-Wigner transformation, 
since $c_{j}^{\pm}$ are not hermitian adjoint each other \cite{lieb}, but they anticommute 
and are of square zero.
\subsubsection{The new vertex case}

The $\eta_{j}^{\pm}$ operators do have a fermion-like behavior. Here $L_{z} = Q$,
\begin{equation}
\displaystyle
\eta_{j}^{\pm} = e^{i\pi(Q_{1} + \cdots + Q_{j-1})}e^{i{\pi\over 2}Q_{j}}\beta_{j}^{\pm}.
\end{equation}
They also anticommute:
$$\eta_{j}^{\pm}\eta_{i}^{\pm} = -\eta_{i}^{\pm}\eta_{j}^{\pm},$$
\begin{equation}
\{\eta_{j}^{+}, \eta_{i}^{-}\} = \delta_{ij}e^{2i\pi(Q_{1} + \cdots + 
Q_{j-1})}\frac{e^{2i\pi Q_{j}} - 1}{-2}.
\end{equation}

Now the r.h.s of the second equation of (40) must be applied to the 
N-fold tensor product states of the type 
\begin{equation}
\displaystyle
\otimes_{l=1}^{N} \vert n_{l} + 1/2\rangle \hspace{1cm} n_{l} = 0, 1, 2, \cdots.
\end{equation}
Since all $Q_{j}$ are diagonal in this representation, we obtain a factor
$(-1)^{j-1}$ as in the second equation of (37):
\begin{equation}
e^{2i\pi\{(n_{1} + \cdots + n_{j-1}) +{(j-1)\over 2}\}}\frac{e^{2i\pi(n_{j} + 
{1\over 2})} - 1}{-2} = (-1)^{j-1}.
\end{equation}

Hence in analogy with the six-vertex case, we may define the fermion-like operators:
\begin{equation}
d_{j}^{\pm} = (i)^{j-1}\eta_{j}^{\pm},
\end{equation}
which anticommute properly as $c_{j}^{\pm}$ in (37), but its square 
(or any power of it), is not zero because:
\begin{equation}
(a_{j}^{+})^{m} = e^{-i{\pi\over 4}m(m-1)}e^{i{\pi\over 2}mQ_{j}}(\beta_{j}^{+})^{m},
\end{equation} 
is not zero at site $j$. This feature already manifests itself in the 
Bethe-Ansatz wave function which 
differs markedly with the Bethe-Ansatz wave function in the six-vertex model.
However since the fermionic character prevails at the point $q = i\; (\eta = {\pi\over 2})$, 
there is a simplifying decoupling in the Bethe-Ansatz equations and perhaps, 
a simple direct evaluation of the free energy is possible.
In any case, this study suggests that the old duality in one dimension between 
Fermions and Bosons discovered in the continuum, has an unexpected features 
on the lattice which are brought to light by the algebraic structure of quantum groups.

\section{Conclusion and outlook}

We have presented here a study of an integrable vertex system made of 
Ising variables on horizontal bonds and, ``oscillator-like'' variables 
on vertical bonds. The model is integrable in the sense it fulfills the Yang-Baxter equations 
with an $R$-matrix identical to the six-vertex model one. This fact leads to 
the same critical behavior of the free energy per site as in the six vertex case \cite{baxlivre}
. The detailed algebraic structure is however different. The Bethe Ansatz wave 
function is not in general antisymmetric and the free fermion limit reveals 
a fermionic behavior of the creation/annihilation operators for the excitations.\\

In the past, there has been several studies of this type of the vertex. 
Faddeev et al.,  Korepin et al. \cite{fada}, \cite{kora}, \cite{fadb}, 
in the Sine-Gordon model on a lattice (or its non relativistic 
limit the non-linear Sch\"odinger model), have considered an integrablity based 
on two sites in order to achieve solubility of the model. Zhou \cite{zhou} has 
performed the Holstein-Primakoff transformation  on the six-vertex model. 
Here we have directly obtained the ``vertical Bose variables'' by the technique 
of quantum groups, and have obtained an integrability based only on one site.\\

The present study opens up new lines of investigation. First it would be 
interesting to find the Hamiltonian of the system, since the Hamiltonian 
of the six-vertex model is the XXZ chain. Presumably, such a Hamiltonian 
under appropriate continuum limit may yield the Sine-Gordon Hamiltonian. 
Second, the limit of free Fermion may provide also new insight in the old 
equivalence between Fermion and Boson in one dimension. Finally one may choose 
formal perturbation theory to generate, as it is known in the continuum, 
a classical Coulomb gaz of ``electric charges'' in the plane. But on a lattice, 
the spacing between sites would then be a parameter controlling the divergences 
which one has to introduce by hand in a standard bosonisation procedure. 
We hope to tackle these problems in the future.\\

\medskip
\noindent{\bf Acknowledgment}

One of us (S.C.L) would like to thank the ``Conseil R\'egional de la R\'egion 
Centre'' for financial support.   

We thank the referee for his comments leading to the improvement
of the paper.

\end{document}